\documentclass[usenatbib]{mnras}

\usepackage{fixltx2e} 

\makeatletter
\newcommand{\textsubsuperscript}[2]{%
  \begingroup
    \settowidth{\@tempdima}{\textsubscript{#1}}%
    \settowidth{\@tempdimb}{\textsuperscript{#2}}%
    \ifdim\@tempdima<\@tempdimb
      \setlength{\@tempdima}{\@tempdimb}%
    \fi
    \makebox[\@tempdima][l]{%
      \rlap{\textsubscript{#1}}\textsuperscript{#2}}%
  \endgroup}
\makeatother

\usepackage{float}
\usepackage{amsmath}
\usepackage{amsfonts}
\usepackage{amssymb}
\usepackage{times}
\usepackage{graphicx}
\usepackage[T1]{fontenc}
\usepackage{soul}
\usepackage{aecompl}
\usepackage{verbatim}
\title[Effects of Metallicity on HMXB Formation]{Effects of Metallicity on High Mass X-ray Binary Formation}
\author[S. Ponnada, M. Brorby, P. Kaaret]{S. Ponnada$^{1}$\thanks{E-mail:
sam-ponnada@uiowa.edu},  M. Brorby$^{1}$\thanks{E-mail:
matthew-brorby@uiowa.edu},  P. Kaaret$^{1}$\\
$^{1}$Department of Physics and Astronomy, University of Iowa, Iowa City, IA 52242\\
}
\begin{document}

\pagerange{\pageref{firstpage}--\pageref{lastpage}} \pubyear{2018}

\maketitle

\label{firstpage}

\begin{abstract}
The heating of the intergalactic medium in the early, metal-poor Universe may have been partly due to radiation from high mass X-ray binaries (HMXBs). Previous investigations on the effect of metallicity have used galaxies of different types. To isolate the effects of metallicity on the production of HMXBs, we study a sample consisting only of 46 blue compact dwarf galaxies (BCDs) covering metallicity in the range 12+log(O/H) of 7.15 to 8.66. To test the hypothesis of metallicity dependence in the X-ray luminosity function (XLF), we fix the XLF form to that found for near-solar metallicity galaxies and use a Bayesian method to constrain the XLF normalization as a function of star formation rate (SFR) for three different metallicity ranges in our sample. We find an increase by a factor of 4.45 $\pm$ 2.04 in the XLF normalization between the metallicity ranges 7.1-7.7 and 8.2-8.66 at a statistical significance of 99.79 per cent. Our results suggest that HMXB production is enhanced at low metallicity, and consequently that HMXBs may have contributed significantly to the reheating of the early Universe.
\end{abstract}

\begin{keywords}
galaxies: dwarf galaxies, starburst --- X-rays: galaxies
\end{keywords}

\section{Introduction}\label{sect:intro}
Around 300,000 years after the Big Bang, the mostly hydrogen plasma that permeated all of space had cooled enough to combine and form neutral atoms. As neutral matter fell into the primordial dark matter gravitational wells, the first stars and galaxies formed and began to heat and ionize the surrounding baryonic matter \protect\citep{Barkana2001}. \protect\citet{Mirabel2011} proposed that high-mass X-ray binaries (HMXBs) in star-forming galaxies were the main source of X-rays in the early universe. Early X-ray binaries heated the intergalactic medium, producing a strong impact on the 21-cm hydrogen recombination signals that serve as primary diagnostics of the physics of the epoch of reionization, as well as impacting the formation and structure of galaxies in the early universe \protect\citep{Artale2015,Fialkov2017,Madau2017}. 

HMXBs have been found to have populations which depend on the star formation rate (SFR) \protect\citep{Griffiths1990,David1992,Ranalli2003,Grimm2003,Kaaret2008,Mineo2012a}, but these samples consist of well-evolved galaxies with near-solar metallicities. The stars and galaxies during the time of reionization had low metallicities. However, direct studies of the sources that produced X-rays in the early universe are currently impractical due to their high redshifts. \protect\cite{Kunth2000} have suggested that blue compact dwarf (BCD) galaxies may be local analogs of the X-ray producing, metal-deficient galaxies found in the early universe. 

Models \protect\cite[e.g.,][]{Dray2007} and studies \citep{Mapelli2010,Kaaret2011,Prestwich2013,Brorby2014} have shown enhanced X-ray binary formation in these low metallicity galaxies. There is now growing evidence which shows that the production of X-ray sources, HMXBs in particular, increases as metallicity decreases \protect\cite[e.g.,][]{Mapelli2010,Kaaret2011,Prestwich2013,Basu-Zych2013,Brorby2014,Brorby2016, Douna2015}. Given that galaxies at very high redshifts have lower metal content, one would expect a larger number of HMXBs in early galaxies than would be predicted using the relation found for nearby, near-solar metallicity galaxies. 
\raggedbottom

Using a sample consisting predominantly of BCDs, \protect\cite{Prestwich2013} showed that the number of the brightest HMXBs, or ultraluminous X-ray sources (ULX), in star-forming galaxies is enhanced for these extremely metal-poor galaxies by a factor of 7 $\pm$ 3. In a comparative sample of near and sub-solar metallicity galaxies, all of which were spiral galaxies, they found no significant increase in the number of ULXs relative to SFR. A similar study by \protect\cite{Douna2015} also measured HMXB populations across a large metallicity range and found enhancement at lower metallicities, but they too used different galaxy types at low and high metallicities (BCDs and spirals, respectively). For a review, see \protect\citet{Kaaret2017}.

These observational results match predictions from simulations done by \protect\cite{Linden2010} who showed a dramatic increase in bright HMXBs below 20 percent solar metallicity.
\protect\cite{Linden2010} explain that this bright HMXB population increase is due to a relative enhancement in the number of binaries accreting through Roche lobe overflow (RLO) rather than wind accretion systems. RLO is the dominant accretion mechanism for close binaries and produces brighter HMXBs than wind accretion, which is more common in binaries with longer orbital periods. A separate set of simulations, done by \protect\cite{Fragos2013a}, showed that the $L_X$/SFR increased by an order of magnitude going from solar metallicity to 10 percent the solar value.

Although many attribute the evolution of the X-ray luminosity function (XLF) to metallicity effects, there may also be effects correlated to the total stellar mass of the galaxy, star formation history, or gas mass fraction. 
However, no studies have attempted to test the evolution of the total luminosity and number of HMXB across a broad range of metallicity while maintaining the uniformity of a single galaxy type. Here, we use a single galaxy type (BCDs) across a large range of metallicities to test the metallicity dependence more strictly by eliminating most of these other variables. We study a large sample of BCDs, varying in metallicity by almost two orders of magnitude, to test for metallicity dependence of the XLF and $L_X$/SFR relations. In Section 2, we discuss our sample selection criteria. Section 3 describes our X-ray image analysis methods, and Section 4 details our procedures for analyzing UV and IR images to determine star formation rates. In Section 5, we discuss the Bayesian method we use for testing XLF evolution over metallicity, our analysis of HMXB formation evolution, and the evolution of $L_X$/SFR over metallicity. We conclude with a summary and discussion in Section 6.

\section{The Sample}
Defined by optically blue continuua in their spectra, BCDs exhibit strong recent star formation, indicating similar star formation histories. Their recent star formation activity would also indicate that their X-ray binary populations are dominated by HMXBs \protect\citep{Colbert2004,Fragos2013a}. Being dwarf galaxies, they also have low stellar masses, with the originally defined magnitude cut-off being M$_{B}$ $>$ -18, thereby providing the expectation that BCDs do not range drastically in mass \protect\citep{Thuan1981}.  The gas mass fraction of BCDs is typically higher than 30 per cent \protect\citep{Zhao2013}; however, \protect\cite{Thuan2016} found that the gas mass fraction appears to decrease with metallicity. \protect\citeauthor{Thuan2016}'s findings should be taken into consideration for our sample, however, we note that their sample did not consist of strictly BCDs.

We expand a set of 25 BCDs studied in \cite{Brorby2014}, hereafter B14, by selecting BCDs over all metallicities up to a distance of 60 Mpc. Cross-referencing the BCD population defined by \cite{Izotov2007} and the NED catalogue with the \textit{Chandra} archive, we find a total of 21 additional BCDs with \textit{Chandra} observations, \textit{GALEX} observations, and published metallicities. We add this to our sample for an overall sample size of 46 BCDs.



For the final sample, we determine star formation rates using \textit{GALEX} (UV) data following the methods used in previous studies, B14 and \citet{Brorby2016}, hereafter B16. We obtain metallicities from published values, all of them obtained through the direct-temperature method, with an exception.  The distance and metallicity measurements from reported values in the literature are listed in Table \ref{tab:sample}.

\section{Analysis}\label{sect:Analysis}\label{subsect:observations}
Following B14, we analyze observations carried out with ACIS-S3, one of the back-illuminated chips aboard \textit{Chandra}, where target galaxies are close to the chip's aimpoint. Reprocessing level 1 event files using the latest versions of CIAO (4.9) and CALDB (4.7.7), we locate X-ray sources within the 0.5$-$8 keV band using the CIAO tool \texttt{wavdetect}. Using a Python 2.7 script and the \texttt{ciao.contrib.runtool}, we automate the data pipeline process. Running the CIAO tool \texttt{mkpsfmap} with an enclosed counts fraction (ECF) of 90 per cent, we use the resulting PSF map for the \texttt{wavdetect} tool, with pixel scales set to the $\sqrt{2}$ series from 1 to 8. The significance threshold (\texttt{sigthresh}) is set to $10^{-6}$, resulting in less than one false detection per image. We use \texttt{maxiter} = 10, \texttt{iterstop} $= 0.00001$, and \texttt{bkgsigthresh} = 0.0001. 

	The source list generated by \texttt{wavdetect} is used in the background-deflaring procedure using the CIAO tool \texttt{deflare}, which first excludes source regions and proceeds to scan the event file for time periods where the background is at the mean count value. From this, good time intervals (GTIs) are extracted to be used in filtering the event file when calculating fluxes. 
	
	To determine whether the sources found using \texttt{wavdetect} are within the target galaxy, we construct $D_{25}$ ellipses using positions and dimensions in the HyperLeda\footnote{\url{http://leda.univ-lyon1.fr/}} data base and the CIAO tool \texttt{dmmakereg}, which takes position parameters and dimensions to create region files. Before creating the region file, we increase the dimensions of our ellipses to cover an extra 200 pc to account for binaries moving from their formation regions, and an additional 1.18 arcsec to account for positional error from HyperLeda and \textit{Chandra}'s astrometric error -- following the procedure of B14. 
	
		
	 We calculate fluxes using the CIAO tool \texttt{srcflux}, which converts source counts to fluxes and uses point spread function models to correct source and background regions. The input parameters for this tool are the source list, the neutral hydrogen column density (found using \texttt{Colden}\footnote{\url{http://cxc.harvard.edu/toolkit/colden.jsp}}), the photon index ($\Gamma = 1.7$), point spread function method ('arfcorr'), and the energy range ($0.5 - 8$ keV). Using \texttt{srcflux}, we calculate unabsorbed fluxes for the given energy interval with an absorbed power law model. The fluxes are later converted to luminosities using the distances to the target galaxies. X-ray source photometry information is listed in Table \ref{tab:Xray}.
	 
We follow the procedure of B14, to determine the minimum number of counts in the 0.5-8 keV range for a 95 per cent probability threshold of source detection. We refer to this as the completeness number. From this, we find completeness luminosities which are the luminosities corresponding to the completeness number for the aforementioned likelihood of source detection. This involves using relations given in \protect\citet{Zezas2007} for the probability of detecting a source with a certain number of counts over a background with a given number of background counts per pixel. From this calculation, we find that the completeness number is 9. Using this minimum number, we utilize the \texttt{ChandraPIMMS}\footnote{\url{http://cxc.harvard.edu/toolkit/pimms.jsp}} tool and $\log N - \log S$ curves of \protect\citet{Georgakakis2008} to determine the minimum fluxes for source detection and the expected number of background sources respectively. We note that in Equation 2 of \protect\citet{Georgakakis2008}, the terms with a coefficient of K$^{'}$ must be adjusted to have minus signs. We convert these minimum fluxes to 'completeness' luminosities, and hereafter refer to the completeness luminosity as L$_\text{min}$.


\section{Star Formation Rates from \textit{GALEX} and \textit{Spitzer}}

	To determine star-formation rates with UV measurements, we obtain images corresponding to each galaxy from the \textit{GALEX}{\footnote{\href{http://galexgi.gsfc.nasa.gov/docs/galex/FAQ/counts_background.html}{\nolinkurl{http://galexgi.gsfc.nasa.gov/}}}} archive in the far-ultraviolet (FUV) and near-ultraviolet (NUV). To find the respective SFRs, we follow the steps taken in B14 and B16, first extracting count rates for each galaxy. By using the CIAO tool \texttt{dmmakereg}, we use our previous method of creating $D_{25}$ ellipses and adjust position and angles individually for the \textit{GALEX} images, using them as source regions. We extract background-subtracted FUV and NUV count rates for the galaxies by using source-centered annular ellipses with areas eight times that of the respective $D_{25}$ region and the CIAO tool \texttt{dmstat}. We then follow the relations given in B14 between count rates, fluxes, and luminosities to subsequently determine the star-formation rates in each UV band. 
	



As the SFR determined by the NUV component does not account for reprocessing of light by dust, we also determine IR components of the SFR for each galaxy using \textit{Spitzer} archive\footnote{\url{http://sha.ipac.caltech.edu/applications/Spitzer/SHA/}} images and following the procedure of B14. 




In Section 3.3.3 of B14, a correlation was found between the FUV SFR measure of \protect\citet{Hunter2010} and the NUV+IR method used in \protect\citet{Mineo2012a} with a slope of 1.23 $\pm$ 0.11. The correlation included galaxies with FUV SFRs up to 0.1 \text{$M_\odot \text{yr}^{-1}$}. 

\begin{figure}
\includegraphics[width=0.48\textwidth]{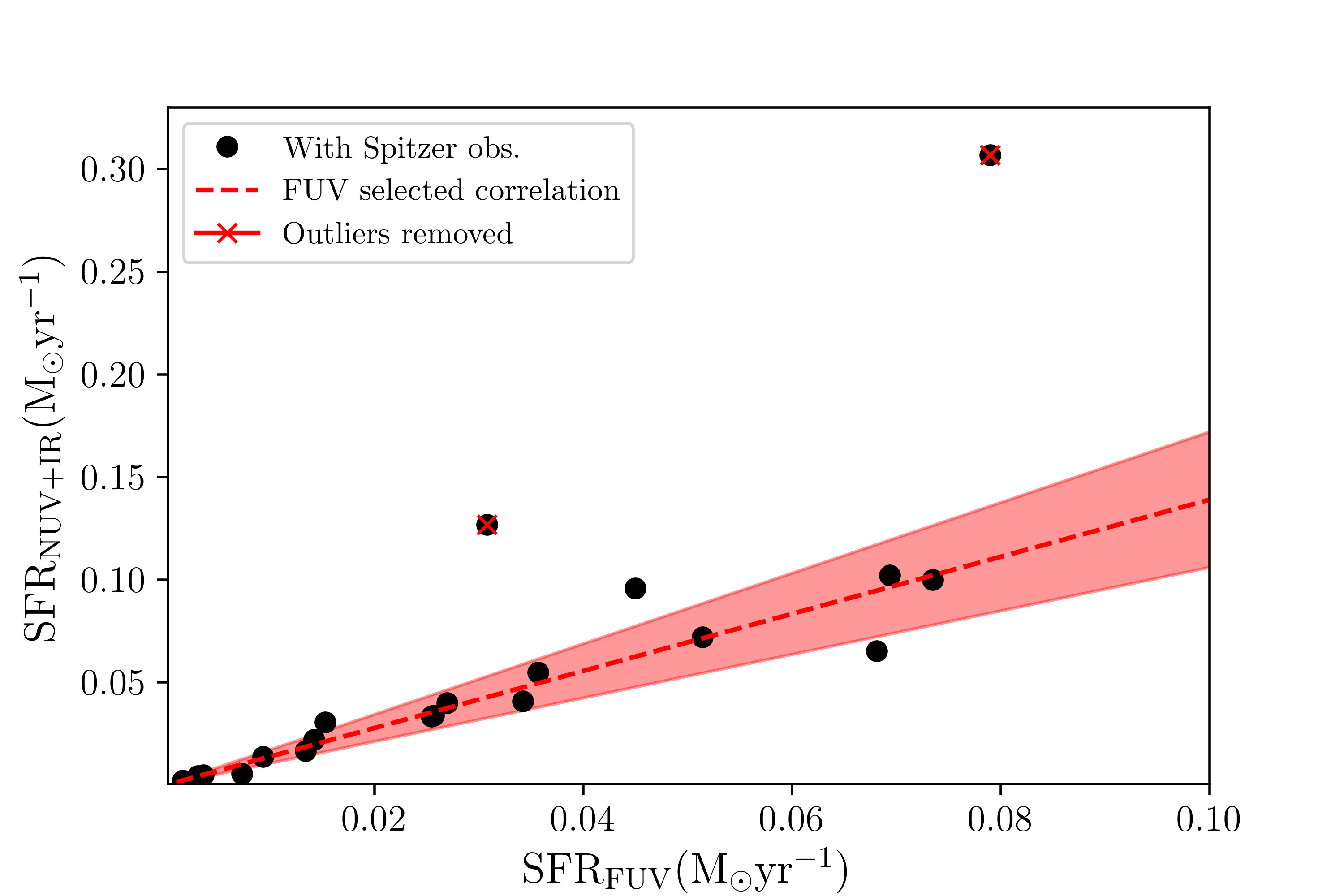}
\caption{\text{SFR$_{NUV+IR}$} vs \text{SFR$_{FUV}$} for the sample galaxies with Spitzer images (black circles). The correlation found for galaxies with \text{SFR$_{FUV}$} < 0.1 (red line) has a slope of 1.39 $\pm$ 0.33. Shaded regions represent uncertainties in slope values.}\label{fig:SFRreanalysed}
\end{figure}

We analyze this correlation between the two SFR methods for our sample.
We find that all of the galaxies lacking IR observations lie within a FUV SFR of 0.1 \text{$M_\odot \text{yr}^{-1}$}. Using 0.1 as a cutoff FUV star formation rate for the linear regression for the galaxies with IR observations, we use this new correlation to determine NUV+IR SFRs for the remaining 18 galaxies. We remove two galaxies, Mrk 996 and NGC 5253, from the regression as they were clear outliers to the observed trend. The new correlation, plotted in red in Figure \ref{fig:SFRreanalysed}, has a slope of 1.39 $\pm$ 0.33, which is consistent with the slope of 1.23 $\pm$ 0.11 found in B14. Our UV and IR measurements and corresponding SFRs are detailed in Table \ref{tab:sfr}.

\raggedbottom

\section{Results}\label{sect::Results}

\subsection{XLF Normalization}\label{subsect::XLF}
For this study, we aimed to determine how the XLF changed with metallicity. In order to do so, we binned our galaxies by metallicity, with a cutoff of $12+\log_{10}({\rm O/H})$ = 7.7 for low metallicity, 7.7 < $12+\log_{10}({\rm O/H})$ < 8.2 for intermediate metallicity, and $12+\log_{10}({\rm O/H})$ $\geq$ 8.2 for high metallicity. Our limit for low metallicity is consistent within rounding for the limit of 7.65 used by \citet{Prestwich2013} for extremely metal-poor galaxies. We chose to bin the intermediate and high metallicity galaxies with the specified cutoff values to maintain a similar number of galaxies with source detections as the low metallicity bin. The bins and their compositions are detailed in Table \ref{tab:binsxlf}.

To determine XLF normalization evolution with metallicity, we employ Bayesian methods described in B14. The XLF has the form
\begin{equation}
\frac{dN}{dL_{38}} = q s L\textsubsuperscript{38}{-$\alpha$}
\end{equation}
where q is the normalization constant, s is the SFR (M$_\odot$ yr$^{-1}$), $L_{38}$ is the $L_X$/10$^{38}$, and $\alpha$ is the power-law index \citep{Grimm2003}.
We take the gamma distribution as the conjugate prior of the Poisson distribution. The gamma distribution is defined by \begin{equation}\label{gammaDist}
\text{GAMMA}(q;X,B) = \frac{B^{X} q^X e^{-qB}}{(X-1)!},
\end{equation}
where $X = \sum x_i$, $x_i$ is the number of observed sources in the $i^\text{th}$ galaxy, and $B = \sum \frac{s_i}{1-\alpha}\left(L_\text{cut}^{1-\alpha} - L_\text{min,\ i}^{1-\alpha}\right)$, where $s_i$ is the SFR (M$_\odot$ yr$^{-1}$) of the i$^{th}$ galaxy. The gamma distribution is the posterior probability distribution, \text{$f^\prime(q|D)$}, for the XLF normalization q given the data D, where the likelihood function is given by the Poisson distribution \textit{P(D|q)}. L$_\text{cut}$ is the cut-off luminosity for the XLF, 110 $\times$ 10$^{38}$ erg s$^{-1}$ \protect\citep{Mineo2012a}. D represents the number of sources, and the minimum luminosity values and SFRs for the binned galaxies are hyperparameters of the prior.

As in B14, GAMMA(q;0,0) is initially taken as a placeholder since it is the "know-nothing" prior. The case of choosing a uniform prior could bias our result as this would be proportional to a prior of GAMMA(q;1,0), which is equivalent to adding a galaxy with a SFR of 0 containing an X-ray source. To this end, we simulate a random sample of 46 galaxies over various SFR ranges and perform the Bayesian calculation using a uniform prior as well as the "know-nothing" prior. We find that at low SFRs, the "know-nothing" prior better matches the data, and high SFRs, the results of both calculations converge as expected due to more data.

In this method, the XLF normalization parameter and its error are given by the mean (\textit{X/B}) and standard deviation $\sqrt{(\textit{$X/B^{2}$})}$ of the gamma distribution. These values are calculated for each metallicity range in the final sample, which is described in Table \ref{tab:binsxlf}. The quoted uncertainties do not take into account uncertainty on star formation rate. 

\begin{figure}
\centering
\includegraphics[width=0.5\textwidth]{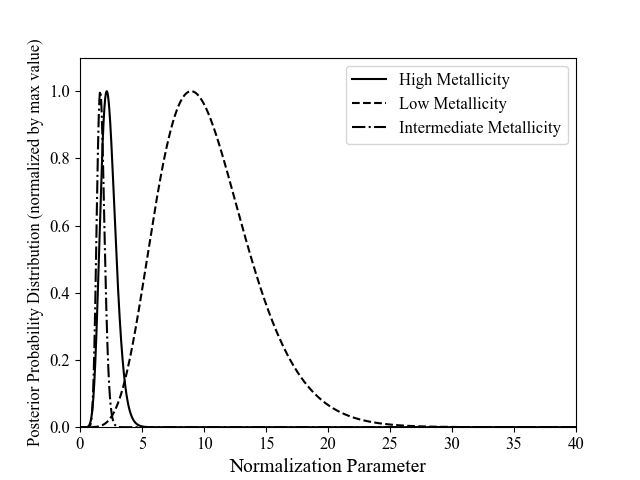}
\caption{Normalization parameters of high metallicity BCDs (solid line), intermediate metallicity BCDs (dot-dashed line), and low metallicity BCDs (dashed line).}\label{fig:xlf}
\end{figure}

Our results for the posterior probability distributions for each metallicity bin are shown in Figure \ref{fig:xlf}. We see a significant increase in the XLF normalization of the low metallicity range, which has a mean metallicity of $12+\log_{10}({\rm O/H})$ = 7.48, relative to the high metallicity ranges, which have mean metallicities of $12+\log_{10}({\rm O/H})$ = 7.97 and 8.37 respectively, indicating 
evolution with metallicity. Our XLF normalization value for the low metallicity bin of 10.38 $\pm$ 3.83 also is considerably increased with respect to the \protect\citet{Mineo2012a} value of 1.49 $\pm$ 0.07 for near-solar metallicity galaxies.

We analyze the significance of these differences by simulating a distribution of a random variable defined as the difference between two random variables drawn from the two gamma distributions, i.e., we test the null hypothesis of the higher metallicity bin having a greater "q" value than the low-metallicity bin. To this end, the test statistic is constructed by subtracting the higher metallicity parameter from the low-metallicity parameter. Since we are testing only whether the higher metallicity value is greater, we opt for the one-sided test. The fraction of values that fall below zero corresponds to the probability that the difference between the two normalization parameters is less than zero. We find the probability of q$_{int}$ and q$_{high}$, the normalization parameters of the intermediate and high-metallicity bins, being greater than the low-metallicty bin to be true at confidences of 2.2 $\times$ 10$^{-4}$ and 2.1 $\times$ 10$^{-3}$ respectively. Alternatively, this can be stated as a 99.98 and 99.79 per cent confidence that the null hypotheses can be rejected. Our normalization value for the low metallicity galaxies is greater than the value found by \protect\citeauthor{Mineo2012a} for near-solar metallicity galaxies at a confidence of 99.994 per cent. 

The normalization parameters for the intermediate and high metallicity bins do not differ from the Mineo value significantly. This suggests that HMXB formation is not influenced by the nature of the host galaxy. Futhermore, \citet{Prestwich2013} found a break in the number of ULXs normalized to SFR and \citet{Douna2015} found a break in the number of HMXBs normalized to SFR and in the XLF at $12+\log_{10}({\rm O/H})$ of $\approx$ 8.0. In conjunction with our result, this indicates that the XLF enhancement seems to diminish in the metallicity range 7.7-8.0.

\begin{table*}
\caption{XLF Normalizations of the metallicity bins}\label{tab:binsxlf}
\begin{tabular}{ccccc}
\hline
\hline\noalign{\vspace{1mm}}
Bin Number & $12+\log_{10}({\rm O/H})$ & Number of galaxies with detections  &  Number of non-detections & XLF Normalization \textit{q} ($M_\odot^{-1} \text{yr}$)\\
1 & 7.1-7.7 & 8  & 18 & 10.38 $\pm$ 3.83 \\
2 & 7.7-8.2 & 8  & 6 & 1.69 $\pm$ 0.32 \\
3 & 8.2-8.66 & 5  & 1 & 2.33 $\pm$ 0.63 \\
\hline
\end{tabular}
\\
\end{table*}

\subsection{$L_X$/SFR - Metallicity relation}\label{subsect::reg}

	In addition to studying XLF normalization, we analyze the relationship between the total HMXB luminosity, SFR, and metallicity. \citet{Douna2015} and B16 found correlations between $L_X$/SFR and metallicity with galaxy samples containing various types over a similar range of metallicity to ours. However, these studies ignored upper-limits in the linear regression procedure.
	
	To take the upper-limits in our sample into account, we follow survival analysis techniques described in \protect\citet{Schmitt1985,Isobe1986}. The method we employ is the Buckley-James regression method, which is conveniently packaged in the Space Telescope Science Data Analysis System (STSDAS). The Buckley-James method uses the non-parametric Kaplan-Meier estimator to modify the regression in the presence of upper-limits by giving large residuals lower weighting \protect\citep{Isobe1986}. Applied statistics literature on the Buckley-James method suggests that the regression method should not be used in cases with over 20 per cent censoring. If we include all the upper-limits in our sample, our censoring fraction would exceed 50 per cent. To see how the regression results change with the number of upper-limits included, we perform the analysis three different ways, including only the 4 upper-limits which constrain the \protect\citet{Mineo2012a} $L_X$-SFR relation, including only the 16 upper-limits which constrain the B16 $L_X$-SFR-metallicity relation, and including no upper-limits. We run the \texttt{buckleyjames} routine within STSDAS with default parameters. 
    
\begin{figure}
\centering
\includegraphics[width=0.5\textwidth]{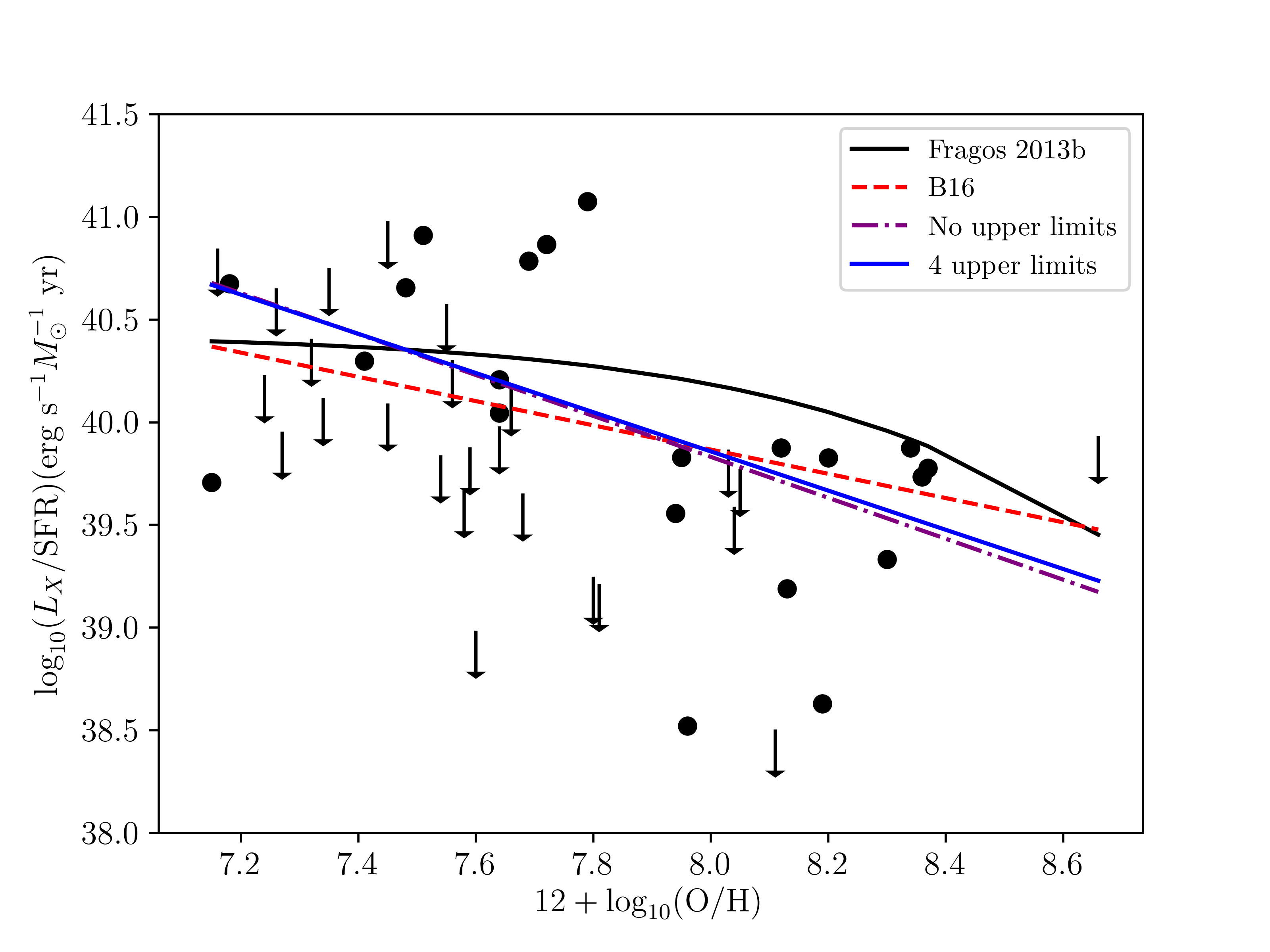}
\caption{L$_{X}$/SFR - metallicity relation for our sample. The black points represent the total source luminosity normalized to SFR for a given galaxy while upper limits given by arrows represent completeness luminosities. The solid, black line represents the model derived from stellar population synthesis simulations of \protect\citet{Fragos2013a}, and the red, dashed line represents the multi-sample correlation found in B16. Our survival analysis correlation for the BCDs is represented by the solid blue line for the 3 upper-limit case. The 0 upper limit case is shown with the purple dot-dashed line. We list regression coefficients in Table \ref{tab:lxsfr}.}\label{fig:lxsfr}
\end{figure}

    We apply this analysis to our sample by fitting an equation of the form 
\[
\text{log($\frac{L_X}{ergs^{-1}}$)} = a \text{log($\frac{SFR}{M_{\odot} yr^{-1}}$)} + b  \text{log($\frac{(O/H)}{(O/H)_{\odot}}$)}  + c
\]
with \textit{a} set to unity. We plot this relation in Figure \ref{fig:lxsfr} along with B16's findings and the relevant model found in \citet{Fragos2013a}. The survival analysis results are tabulated in Table \ref{tab:lxsfr}. 


 Our measured slopes and their uncertainties vary as a function of the number of upper-limits included. When including the 16 upper-limits that constrain the B16 relation, the survival analysis seems to break down due to the abundance of censored data points. In the low-censoring case, including only the 4 constraining \protect\citeauthor{Mineo2012a}, the result indicates a negative correlation between the total resolved point source luminosity normalized to SFR and metallicity with slope of -0.954 $\pm$ 0.37, consistent with \protect\citeauthor{Douna2015}'s value of -1.01 and B16's value of -0.59 $\pm$ 0.13. Analysing the trend when excluding all upper-limits, as done in \protect\citeauthor{Douna2015} and B16, yields similar results to those studies, with an observed slope of -0.996 $\pm$ 0.38. 


\def\arraystretch{1.5}
\begin{table*}
\caption{L$_{X}$/SFR - metallicity regression coefficients}\label{tab:lxsfr}
\begin{tabular}{lccc}
\hline
\hline\noalign{\vspace{1mm}}
Number of upper-limits & Slope & Intercept & Standard deviation of regression \\
0 & -0.996 $\pm$ 0.38 & 39.14 $\pm$ 0.34 & 0.359 \\
4 & -0.954 $\pm$ 0.37 & 39.20  & 0.654 \\
16 & -0.446 $\pm$ 0.38  & 39.28 & 0.649 \\

\hline

\end{tabular}
\\
\raggedright\textbf{Notes.}
STSDAS results for Buckley-James regression in the case of censoring and linear regression results for 0 censoring case.
\end{table*}

\section{Summary and Discussion}
Previous studies have shown enhanced production of HMXBs in samples of low metallicity galaxies \protect\cite[e.g.,][]{Mapelli2010,Kaaret2011,Prestwich2013}, as well as B14, however none have studied the evolution of HMXB populations with a galaxy sample of uniform type over a broad range of metal content. 
From our study of catalogued BCDs over a wide metallicity range, we observe a significant increase in the XLF normalization at low metallicity relative to high by a factor of 4.45 $\pm$ 2.04 at a statistical significance of 99.79 per cent, with no significant distinction between high and intermediate metallicity. Future studies on this subject may choose to focus on a large sample of galaxies of the same type with $12+\log_{10}({\rm O/H})$ between 7.7 and 8.0 to elucidate the mechanism behind the observed decrease in the XLF. To probe potential changes in the XLF shape parameter due to variations in the dominant accretion mechanism, a larger sample of HMXB detections, especially at lower metallicities, would also be required.


By utilizing survival analysis regression methods, we build upon previous correlations found between $L_X$/SFR and metallicity for different censoring compositions. We find agreement with previous relations in the low-censoring and no censoring cases, and unconstraining results in the high-censoring case.

Eliminating potentially confounding variables due to galaxy type in our study of solely BCDs, we find enhanced production of HMXBs at low metallicity and a correlation between $L_X$/SFR and decreasing metallicity. Our results further indicate that in the low-metallicity environment of the early Universe, the production of HMXBs would have been enhanced and thus HMXBs could be more important than previously thought in determining the thermal conditions of the epoch of reionization. 

While we assume elimination of other factors not discussed in this study such as stellar masses, dust fractions, gas mass fractions, and star formation histories due to maintaining a uniform galaxy type, future studies may probe the effects of these variables on the evolution of HMXBs. Since we cannot disentangle the potentially changing gas mass fraction with metallicity in this study, future studies with gas mass measurements may wish to explore potential correlations with HMXB formation. The results from this study can further be used to evaluate effects on the early Universe 21-cm hydrogen recombination signals.

\section*{ACKNOWLEDGEMENTS}

The authors thank the referee, Dr. Leonardo J. Pellizza, for the constructive feedback provided on the manuscript.
M.B. thanks Hai Fu for insightful discussions which helped to improve the quality of the paper.
STSDAS is a product of the Space Telescope Science Institute, which is operated by AURA for NASA.
The scientific results reported in this article are based on joint observations made by the Chandra X-ray Observatory and the NASA/ESA Hubble Space Telescope. Support for this work was provided by the National Aeronautics and Space Administration through Chandra Award No. AR7-18005X issued by the Chandra X-ray Observatory Center, which is operated by the Smithsonian Astrophysical Observatory for and on behalf of the National Aeronautics and Space Administration under contract NAS8-03060. 


\bibliographystyle{mnras}
\bibliography{MyRefs2}

\appendix
\section{Tables}
\def\arraystretch{1.0}
\begin{table*}
\centering
\caption{ BCDs including the Brorby~et~al.~(2014) and NED sample.}\label{tab:sample}
\begin{tabular}{lrccccc}
\hline
\hline\noalign{\vspace{0.05mm}}
Name         	&ObsID        	 &$12+\log_{10}({\rm O/H})$ &RA  &DEC   &Exposure (ks) &Distance (Mpc)\\
\hline\noalign{\vspace{0.05mm}}
DDO 68  &  11271  &  $7.15^a$  &   09 56 45.7  &   +28 49 33.6  &  10.0  &  5.9 \\
I Zw 18  &  805  &  $7.18^a$  &   09 34 02.4  &   +55 14 26.4  &  40.0  &  18.2 \\
SBS 1129+576  &  11283  &  $7.41^a$  &   11 32 02.42  &   +57 22 45.4  &  14.75  &  26.3 \\
SBS 0940+544  &  11288  &  $7.48^a$  &   09 44 17.15  &   +54 11 28.5  &  16.83  &  22.1 \\
RC2 A1116+51  &  11287  &  $7.51^a$  &   11 19 34.36  &   +51 30 12.3  &  11.65  &  20.8 \\
.. & .. & .. &.. &.. &.. & .. \\
\hline\noalign{\vspace{0.05mm}}
\end{tabular}
\\
\raggedright\textbf{Notes.} The full table is available in the online version. \\
The table includes the name, Chandra ObsID, metallicity, RA and DEC (J2000), Chandra exposure time, and distance for each galaxy. \newline References -- \protect\citet{Brorby2014}$^a$, \protect\citet{Thuan2016}$^b$, \protect\citet{Izotov2014}$^c$,  \protect\citet{Engelbracht2008}$^d$, \protect\citet{Zhao2013}$^e$, \protect\citet{Izotov&Thuan2004}$^f$, \protect\citet{Zhao&Gao&Gu2010}$^g$ , \protect\citet{Guseva2000}$^h$, \protect\citet{Kreckel2015}$^i$, \protect\citet{Pilyugin2007}$^j$.
\end{table*}

\begin{table*}
\centering
\caption{X-Ray Source Photometry}\label{tab:Xray}
\begin{tabular}{l c c l r c c c c}
\hline
\hline\noalign{\vspace{1mm}}
Name	&	${n_H}^a$	&		\multicolumn{2}{c}{${R_{25}}^b$}	&		Angle$^b$		&	${N(>L_\text{min})}^c$	&	${N_\text{bkg}}^d$	& ${L_\text{min}}^e$ \\
		&	$(10^{20} \text{cm}^{-2})$		&	\multicolumn{2}{c}{(arcsec)}	&	(deg)	&	& & $\left(10^{38} \text{erg s}^{-1}\right)$\\
\hline\noalign{\vspace{1mm}}
DDO 68 & 1.97 & 66.49 & 32.94 & 77.7 & 2 & 0.3096 & 0.286 \\
I Zw 18 & 1.99 & 11.16 & 8.86 & -55.0 & 1 & 0.0449 & 0.548 \\
SBS 1129+576 & 0.87 & 23.81 & 9.6 & -72.0 & 1 & 0.0452 & 3.767 \\
SBS 0940+544 & 1.34 & 22.69 & 12.58 & -63.5 & 1 & 0.0614 & 2.354 \\
RC2 A1116+51 & 1.19 & 16.38 & 11.44 & -0.7 & 1 & 0.0308 & 3.003 \\
... & ... & ... &... &... & ... &... &... \\
 \hline\noalign{\vspace{1mm}}
 \end{tabular}

\raggedright\textbf{Notes.}\\
The full table is available in the online version. \\
$^a$ Column densities were found using the \texttt{Colden}\footnotemark[2] tool.\\
$^b$ Properties of the D$_{25}$ ellipses were taken from the HyperLeda\footnotemark[1] database and modified as described in Section~\ref{subsect:observations}. These represent 0.5 $\times$ D$_{25}$ dimensions. \\
$^c$ Number of observed sources within the D$_{25}$ ellipse with luminosity greater than L$_\text{min}$.\\
$^d$ Number of expected background sources within the D$_{25}$ ellipse, as determined by $\log N - \log S$ curves of \protect\citet{Georgakakis2008}.\\
$^e$ Luminosities limits calculated for the $0.5-8.0$ keV energy band.\\
$^*$ Denotes galaxies without source detections constraining the Mineo relation.
\end{table*}

\begin{table*}\scriptsize
\centering
\caption{UV Measurements and SFRs.}\label{tab:sfr}
\begin{tabular}{l c c c c c c c}
\hline
\hline\noalign{\vspace{1mm}}
Name	&	$E(B-V)^a$ & Count Rate$_\text{FUV}^b$ & $L_\text{FUV}^c$ &	$\text{SFR}_{\text{FUV}}^d$	& Count Rate$_\text{NUV}^e$ & $\text{SFR}_{\text{NUV},0}^f$ & $\text{SFR}_{\text{IR}}^f$ \\
& (mag)	& (cps) & $\left(10^{26} \text{erg s}^{-1} \text{Hz}^{-1}\right)$	&	$\left(10^{-3} M_{\odot}\ \text{yr}^{-1}\right)$ & (cps)	& $\left(10^{-3} M_{\odot}\ \text{yr}^{-1}\right)$ & $\left(10^{-3} M_{\odot}\ \text{yr}^{-1}\right)$	\\
\hline\noalign{\vspace{1mm}}
DDO 68 & 0.0158 & 20.572 & 1.055 & 13.395 & 65.883 & 15.686 & 0.884\\
I Zw 18 & 0.0292 & 9.918 & 5.362 & 68.1 & 27.894 & 63.197 & 2.025\\
SBS 1129+576 & 0.0115 & 2.849 & 2.809 & 35.67 & 10.229 & 48.395 & 6.465\\
SBS 0940+544 & 0.0114 & 1.605 & 1.116 & 14.175 & 5.639 & 18.837 & 3.196\\
RC2 A1116+51 & 0.0128 & 4.325 & 2.693 & 34.203 & 13.776 & 40.764 & 0.013\\
... & ... & ... & ... & ... & ... & ... & ... \\
\hline\noalign{\vspace{1mm}}
\end{tabular}
\footnotesize
\\
\raggedright\textbf{Notes.}\\
The full table is available in the online version.\\
$^a$ Extinctions were found using the Infrared Science Archive (IRSA) tool DUST.\\
$^b$ Count rate in the FUV band of GALEX observations.\\
$^c$ Luminosity in the FUV band, $1350-1750$\AA . \\
$^d$ Star-formation rate as determined by the \citet{Hunter2010} method.\\
$^e$ Count Rate in the NUV band of GALEX, $1750-2800$\AA . \\
$^f$ Star-formation rate in the NUV.
$^e$ Star-formation rate in the IR via Spitzer images \citep{Brorby2014}.\\
\end{table*}




\end{document}